\documentclass[namedreferences]{solarphysics}

\usepackage[hyperref,optionalrh,natbib]{spr-sola-addons} % For Solar Physics 
\usepackage{graphicx}        % For eps figures, newer & more powerfull
\usepackage{amssymb}        % useful mathematical symbols
\usepackage[usenames]{color}           % For color text: \color command
\usepackage{breakurl}        % For breaking URLs easily trough lines
\ifx \doiurl    \undefined \def \doiurl#1{\href{http://dx.doi.org/#1}{\textsf{DOI}}}\fi
\ifx \adsurl    \undefined \def \adsurl#1{\href{http://adsabs.harvard.edu/abs/#1}{\textsf{ADS}}}\fi
\ifx \arxivurl  \undefined \def \arxivurl#1{\href{http://arxiv.org/abs/#1}{\textsf{arXiv}}}\fi
\ifx \urlurl    \undefined \def \urlurl#1{\href{http://#1}{\textsf{#1}}}\fi
\ifx \emailurl  \undefined \def \emailurl#1{\\ email: \href{mailto:#1}{\textsf{#1}}}\fi
\newcommand{\edit}[1]{{#1}}

\begin{document}

\begin{article}

\begin{opening}

\title{Application of Mutual Information Methods
in Time--Distance Helioseismology}

\author{Dustin~\surname{Keys}$^{1,2}$\sep
        Shukur~\surname{Kholikov}$^{2}$\sep
        Alexei~A.~\surname{Pevtsov}$^{3}$      
       }
\runningauthor{D. Keys \emph{et al.}}
\runningtitle{Mutual Information in Time--Distance Helioseismology}

\institute{$^{1}$ University of Arizona, Tucson, AZ, 85721, USA, \emailurl{dmkeys@gmail.com} \\
$^{2}$ National Solar Observatory, Tucson, AZ, 85719, USA, \emailurl{skholikov@nso.edu} \\
$^{3}$ National Solar Observatory, Sunspot NM, 88349, USA, \emailurl{apevtsov@nso.edu}
}
\begin{abstract}
    We apply a new technique, the mutual information (\edit{MI}) from information theory,
to time--distance helioseismology, \edit{ and demonstrate that it can successfully reproduce
several} classic results  based on the widely used cross--covariance method. 
MI quantifies the 
deviation of two random variables from complete independence, and represents 
a more general method for detecting dependencies in time series than the
cross--covariance function, which only detects linear relationships.
\edit{ We provide a brief description of MI-based  technique and discuss the 
results of the application of MI to derive the 
solar differential profile, a travel-time deviation map for a sunspot and a 
time--distance diagram from quiet Sun measurements.} 
\end{abstract}
\end{opening}
%-------------------------------------------------

\section{Introduction}
     \label{S-Introduction} 
The field of local helioseismology is primarily concerned with the
propagation of high-degree acoustic waves in the solar medium. For 20
years the primary method for detecting these waves has been the
cross--covariance, or cross-correlation function [CCF: \cite{duvall1993}].
While this tool has been used to great success, it
does have limitations. The CCF can be thought of as a time-displaced
scalar product and in some sense this works for detecting when two signals
are similar to one other. However, \edit{the CCF-based method} is at best a first-order
approximation to quantifying the actual relationship between the
signals. Here we propose an alternative tool based on information theory, specifically
the mutual information (MI), which can be used to produce similar results
to the CCF but allows one to characterize the amount of information transferred
from one place on the Sun to the other, and thus, may represent more physical 
``connection" between two regions.

Since \cite{shannon1948} first formalized the theory of information
for communication theory, it has found many uses in
a wide variety of other fields ranging from biology to economics, or
indeed anywhere stochastic models may be used.  Shannon originally set
out to find a measure \edit{[$H$]} of how much uncertainty is in a random process,
or how much information could be encoded in that process. He required
a few basic properties of this measure
such as continuity and a natural behavior with 
respect to probabilities, \emph{i.e.} in the discrete case if all probabilities
are $1/n$ then $H$ should be a monotonically increasing function of $n$
and if two outcomes should be grouped together then $H$ should be a 
weighted sum of the individual values of $H$ for each layer of grouping.
He showed that the only measure satisfying these properties was
the entropy. He then went on to find the rate of transmission of
information through a channel with noise and bandwidth limitations.
This idea was generalized from the rate of transmission to 
\edit{ mutual information}, which
broadly describes the information or entropy which is shared by two
signals, \textit{i.e.} the amount of uncertainty in one signal that is due to
another signal and \textit{vice versa}. This gives a much more general
view of correlations because it captures more dependencies.

When applied to the field of helioseismology, MI offers a new
perspective on old problems, aside from providing an alternative and
independent method to the CCF.  For instance, by tracking the amount of
information in a wavepacket as it propagates across the surface one can
determine information flows, which \edit{ potentially}  can be used in nonequilibrium
thermodynamics \citep[\emph{e.g.}][]{sagawa2012} and also to determine 
wavepacket lifetimes.  By looking at MI between areas of the Sun, 
one can outline regions of information exchange, areas where the dynamics
in one part of the Sun are influencing the dynamics in other, offering
insight into the degree to which different parts of the solar atmosphere
are connected. In this article, we explore the applicability of MI methods to local
helioseismology on the Sun.  We use MI to reproduce some of the results
of traditional time--distance helioseismology as well as provide insights
on the rate of information loss of waves in the solar atmosphere,
which we then use to estimate wave lifetimes. At each step, comparison
is made between the results of MI and CCF \edit{ which we find to be 
in a good agreement}. The rest of the article is organized as follows: Section
\ref{sec:method} introduces mutual information and its computation,
and provides details of our implementation of MI to solar data. Sections
\ref{sec:data}\,--\,\ref{sec:discuss} describe the data and discuss the
results of our analysis.

\section{\edit{Methodology}}
\label{sec:method}
\subsection{\edit{Definitions and Interpretations of MI}}
Let us consider the time series of an acoustic source on the Sun as
realizations of a random variable $X$, and the time series at a
different point as realizations of a random variable $Y$.  The mutual
information [$I(X;Y)$] between the two random variables is
defined as \citep[\emph{e.g.}][]{coverthomas}
\begin{equation}
I(X;Y) = \int_Y\int_X p(x,y)\log\frac{p(x,y)}{p(x)p(y)}{\mathrm d}x{\mathrm d}y,
\label{midef}
\end{equation}
where the integral is taken over all possible outcomes [$x$ and $y$]
with probability distributions $p(x)$ and $p(y)$, and joint probability
$p(x,y)$. The base of the logarithm defines the units of information,
and in this article we take the bases of all logarthims to be the natural
one, corresponinding to natural units of information [nats]. An
alternate relation describes MI in terms of the entropy of the source
[$H(X)$] and the conditional entropy, \edit{[$H(X\big|Y) = -\int p(x,y)\log 
p(x\big|y)$] of the source when} the other signal is known, or {\it vice versa}

\begin{equation}
I(X;Y) = H(X) - H(X\big|Y) = H(Y) - H(Y\big|X).
\end{equation}

If one thinks of the conditional entropy as the amount of information
left in one signal when the other signal is known, then subtracting it
from the total entropy of the signal leaves the amount of information
which is accounted for by the other signal. \edit{One last useful relation
gives MI in terms of the entropies of the signal and the joint entropy,
[$H(X,Y) = -\int p(x,y)\log p(x,y)$] of the two signals}

\begin{equation}
\edit{I(X;Y) = H(X) + H(Y) - H(X,Y)}.
\label{MIjoint}
\end{equation}
\edit{
If we think of the entropy as the uncertainty in a random variable,
then the MI captures the {\it reduction} in uncertainty of one random 
variable due to the presence of the other. Since the maximum 
uncertainty two variables can have is the sum of the uncertainties
of the two variables (knowing one variable does not make the other more
random), it is easy to see that $I(X;Y) \ge 0$ since $H(X,Y) \le H(X)
+ H(Y)$ with equality only in the case of complete independence.}
\par
\edit{
The manner in which the joint probability distribution is used
instead of its covariance captures more
complicated relationships between the two signals than the linear CCF.
The ratio of the joint probability to the product of the marginal
probabilities is a measure of deviations from complete independence. 
We can see that MI is the expectation of the logarithm 
of this ratio. If two signals are completely independent 
then this ratio is exactly 1 and the MI is zero. To see how the MI
behaves under linear correlations one can consider two correlated Gaussian
random variables with zero mean. The covariance matrix will be
\begin{equation}
	\Sigma = \left(
	\begin{array}{cc}
		\langle X^2\rangle & \langle XY\rangle \\
		\langle XY\rangle & \langle Y^2\rangle \\
	\end{array}
	\right).
\end{equation}
The entropy of a univariate Gaussian is $H(X) = \frac{1}{2}\log(2\pi e 
\langle X^2\rangle)$
and for the multivariate case of $n$ variables with covariance matrix $\Sigma$
is $H(X_1,...,X_n) = \frac{1}{2}\log\left(\left(2\pi e\right)^n\big|\Sigma\big|
\right)$ which means by Equation (\ref{MIjoint}) that the MI for our case of two 
variables is
\begin{eqnarray}
I(X;Y) &=& \frac{1}{2}\log\left(2\pi e \langle X^2\rangle\right) + 
	\frac{1}{2}\log\left(2\pi e \langle Y^2\rangle\right) \nonumber \\
	&\:&\; -\frac{1}{2}\log\left[\left(2\pi e\right)^2\left(\langle 
	X^2\rangle\langle Y^2\rangle -\langle XY\rangle^2\right)\right] \nonumber\\
	&=& -\frac{1}{2}\log\left(1 - \frac{\langle XY\rangle^2}{\langle X^2\rangle
	\langle Y^2\rangle}\right),
\end{eqnarray}
and thus the MI is a simple function of the linear correlation between the 
variables. In this notation we can write the CCF as 
\begin{equation}
C(\tau) = \frac{\langle X_t Y_{t+\tau}\rangle}{\sqrt{\langle X^2\rangle
\langle Y^2\rangle}},
\label{ccfdef}
\end{equation} 
where we've indexed the realizations of the random variable by
time and taken the time average. For weakly correlated Gaussians we might then 
expect that 
\begin{equation}
	I(\tau) \approx C(\tau)^2.
\label{ccfsq}
\end{equation}
}
\subsection{\edit{Calculating MI}}
The simplest algorithms for MI use histograms to calculate
the probability distributions. \edit{ This approach has} the advantage of
speed, \edit{ although it} results in an answer that may depend on the binning procedure
used, as in \inlinecite{leontitsis2001} where \edit{ the histogram method is used to derive}  
the mutual average information. 
In order to accurately calculate the MI between two signals, we turn to an 
approach which uses a $k$-nearest neighbor algorithm for estimating the 
probability distribution \citep{kraskov2004}. This algorithm, as opposed to 
simple nearest neighbor algorithms \citep{kozachenko1987,victor2002}, 
has the advantage that by
choosing $k$ one can tune the amount of systematic error \textit{versus} the
amount of statistical error present in the answer.
\inlinecite{kraskov2004} found empirical scaling laws based on $k/N$ and 
recommend using a $k$ between $2-4$. The disadvantage of  
using MI in time--distance helioseismology is that the signals are
typically shorter than the ideal statistical sample. For quiet-Sun
calculations the problem \edit{ can be} alleviated by stringing                                                     
multiple days together, but for active-Sun calculations such as                                                  
detecting sunspot phase travel time deviations, the amount of data is 
\edit{limited by the amount of time the sunspot is visible.  Nevertheless,}
\edit{ as we demonstrate below}, despite not having an optimal statistical sample  
MI can successfully be used to calculate travel time deviations in
sunspots.

When applying the MI method, we follow a similar recipe as in the CCF--based
time--distance method. First, we choose a point on the Sun as an
acoustic source and measure waves propagating outward from the source
by looking for time sensitive correlations in the signal of the source 
and the averaged signal of an annulus centered at the source.
Ideally, one would compare the source with every point on the annulus
centered at the source, but this would require a much faster computer than
the ones that were available to us at the time of this project.
In helioseismology with the CCF, the signals are compared to time displaced
versions of the other. Here we use the time-displaced signals to construct
a joint signal [$Z$] for which we then find the probability
distributions. A point in $Z$ consists of the Cartesian product of a
point from the source and a point from the annulus. We pair them up
according to the time we are looking at for wave propagation, the
correlations are maximum when the time is exactly the time it takes
for a physical wave to travel the distance between the annulus and the
source. \edit{Rather than use a histogram to get the probability density,
an approach which depends on the binning procedure and may not converge,
we used the algorithm of \citet{kraskov2004}, which uses the probability 
[$P_k(\epsilon)$] that there are $k-1$ neighbors in the neighborhood 
defined by the distance [$\epsilon/2$] to the $k$th neighbor. Using this 
probability it is possible to construct an estimate of the probability mass 
[$\rho_i(\epsilon)$] of the $\epsilon$--neighborhood, which is assumed constant 
throughout the entire $\epsilon$--neighborhood [$\rho_i(\epsilon) \approx 
\epsilon^2 p(Z_i)$ for the joint entropy and $\rho_i(\epsilon) \approx 
\epsilon p(X_i)$ for the marginal entropy]. Using $P_k(\epsilon)$, it can be 
shown that  $\langle \log p_i\rangle = \psi(k)-\psi(N)$, 
where  $\psi(x) = \frac{\mathrm{d}}{\mathrm{d}x}\log\Gamma(x)$ is the digamma 
function, and the problem is then reduced to finding the expectation of 
$\log \epsilon_i$, which will be expressed in terms of the number of points
within 1D Euclidean neighborhoods with radius equal to the $k$th nearest neighbor
distance.} The maximum norm is chosen as
a measure of distance, where the distance between points $(x_1,y_1)$ 
and $(x_2,y_2)$ is given by
\begin{equation}
d = \max\left\{\big|x_2-x_1\big|,\big|y_2-y_1\big|\right\}.
\end{equation}
\edit{The maximum norm is chosen because it has a square 
$\epsilon$--neighborhood, which simplifies the calculation.}
The length of the signal is such that a simple sorting algorithm can be
used, in which the points are sorted in one direction, say $x$, and a
running list of neighbors, sorted under the maximum norm, is kept. The
search stops when the distance of the next point in the $x$-direction
exceeds the distance in the maximum norm of the $k$th nearest
neighbor.  The next part of the calculation involves finding the number
of neighbors [$n_x(x,y)$ and $n_y(x,y)$] within the one-dimensional
Euclidean neighborhoods of $x$ and $y$ defined by the $k$th nearest neighbor
distance [$d_k(x,y)$].  The calculated MI [$\hat{I}(X;Y)$] is then
given by
\begin{equation}
\hat{I}(X;Y) = \psi(k)-\langle \psi(n_x+1)+\psi(n_y+1)\rangle+\psi(N),
\label{micalc}
\end{equation} 
\noindent
where $N$ is the length of the time series, and the brackets denote the
expectation value taken over the entire time series.

\subsection{\edit{Application of MI to Helioseismology}}
When the signal from the point and the signal from the arc are paired
up for different times $\tau$, and the MI [$\hat{I}(\tau)$] is calculated
for the joint signal \edit{ one needs to fit a function to extract meaningful 
information. Based on Equation (\ref{ccfsq}) we adopt a representation of MI
by (approximately) the square of the Gabor 
wavelet currently used for the CCF,}
\begin{equation}
\hat{I}(\tau) \approx A\cos^2\left( 2\pi\nu\left(\tau - \tau_p\right)\right)
\exp\left(\frac{\tau-\tau_g}{2\sigma}\right)^2 + B,
\label{cos2}
\end{equation} 
where $A$ is the amplitude of the wavelet, $\nu$ is the frequency
corresponding to the five-minute oscillations, $\tau_p$ is the phase
time, $\tau_g$ is the group time, $\sigma$ describes the width of the
wavelet, and $B$ is a generally small but non-zero constant describing
the \edit{average contribution from higher order correlations.} 
Once fitted, the parameters can be used to calculate travel-time deviations
as a response to flows, and even the information lost as the wave travels 
and correlations are destroyed. \edit{We suspect that this function reveals
a property of the joint probability distribution in that it is well modeled by
Gaussians, but the actual distribution is more complicated as the value of MI
does not quite equal the square of the CCF. A more rigorous derivation 
would require the analytical form of the joint probability density function 
for each $\tau$. The probability density is easy to work with on computational 
problems but finding an analytical form for it is not a trivial task, and is
beyond the scope of this paper.}

\section{Data}
\label{sec:data}

In this study we use medium \textit{$\ell$} [0\,--\,300] spherical harmonic
(SH) time series from the \textit{Michelson Doppler Imager} 
\citep[MDI,][]{sherrer1995} onboard the
\textit{Solar and Heliospheric Observatory}  \citep[SOHO,][]{sherrer1995}.
SH coefficients for these $\ell$, though they are derived from the full disk,
contain information about localized propagating wavepackets.
Time--distance analyses are based on cross--covariance measurements
between different locations separated by some angular distance on the
solar surface. Acoustic waves with the same horizontal phase speed propagate along
approximately the same raypath in the solar interior
\citep{duvall1993}.  In order to isolate acoustic waves within
a particular wave packet bouncing with a particular travel distance, we
perform phase-speed filtering of the SH coefficients.  This procedure
is well accepted and widely used in local helioseismology.
To obtain filtered velocity images, we chose a specific phase velocity
[$\omega/k_h$], took the product of it with each SH time series in the
Fourier domain, and performed the inverse Fourier transform.  Using
filtered SH coefficients, we reconstructed velocity images containing
waves which propagate to a certain range of distances from a given
location.  Details of such filtering are described by
\citet{kholikovhill2014} and \citet{kholikov2014}.

To demonstrate the sensitivity of the MI technique to acoustic travel-time
perturbations and solar subsurface flows, we performed three types
of measurements:  \edit{a calculation of the solar differential profile, 
a map of sunspot travel-time deviations, and a time--distance diagram in quiet Sun.}

Mean travel times can be measured using the
center-to-annuli scheme. For this purpose an MDI time series on 23 October
2003 was used. Filtered and reconstructed velocity images centered
on AR\,10484 were generated to measure travel times within
and around the active region. Velocity images
 are tracked relative to the noon time according to the solar
 differential-rotation rate.  In order to measure travel-time
differences in the East-West direction, 15 daily velocities were
reconstructed without tracking.  To avoid projection effects due to 
\edit{ the tilt of the ecliptic with respect to the solar equatorial 
plane [or $B_0$--angle]}, several days around the time period when $B_0$ was
close to zero were used.  To produce multi-bounce time--distance
measurements, three consecutive days of quiet regions are used in 1996,
also around the $B_0=0$ time period. These three days of data were tracked
relative to the middle of the time period. The \edit{ details on the data used 
for these three tests} are listed in Table 1.

\begin{table}
\caption{Dates and phase-speed filter parameters of MDI time series used}
\begin{tabular}{ccccc}
\hline
Date & Central $\ell$ & Phase speed [$\mu$Hz $\ell^{-1}$] & $\Delta$ range [deg.] & Used purpose\\\hline
23 Oct. 2003 & 140 & 20.5  & 7-10  & sunspot\\
11\,--\,13 Jun. 1996& 140, 210 & 20.5, 9.5 & 3-40 & TD\\
05\,--\,08 Dec. 1997 & 140 & 20.5 & 7-10 & diff. rotation\\
05\,--\,08 Dec. 1998& 140 & 20.5 & 7-10 & diff. rotation\\
05\,--\,07 Dec. 1999& 140 & 20.5 & 7-10 & diff. rotation\\
07\,--\,10 Dec. 2003& 140 & 20.5 & 7-10 & diff. rotation\\\hline
\end{tabular}
\end{table}       

\section{Results}
\label{sec:results}

In order to show the reliability of the MI method, we reproduce
some of the known results from time--distance helioseismology, which were
based on CCF. 

\subsection{Differential Rotation Profile}
\begin{figure}[t]
\centering
\includegraphics[width=\textwidth]{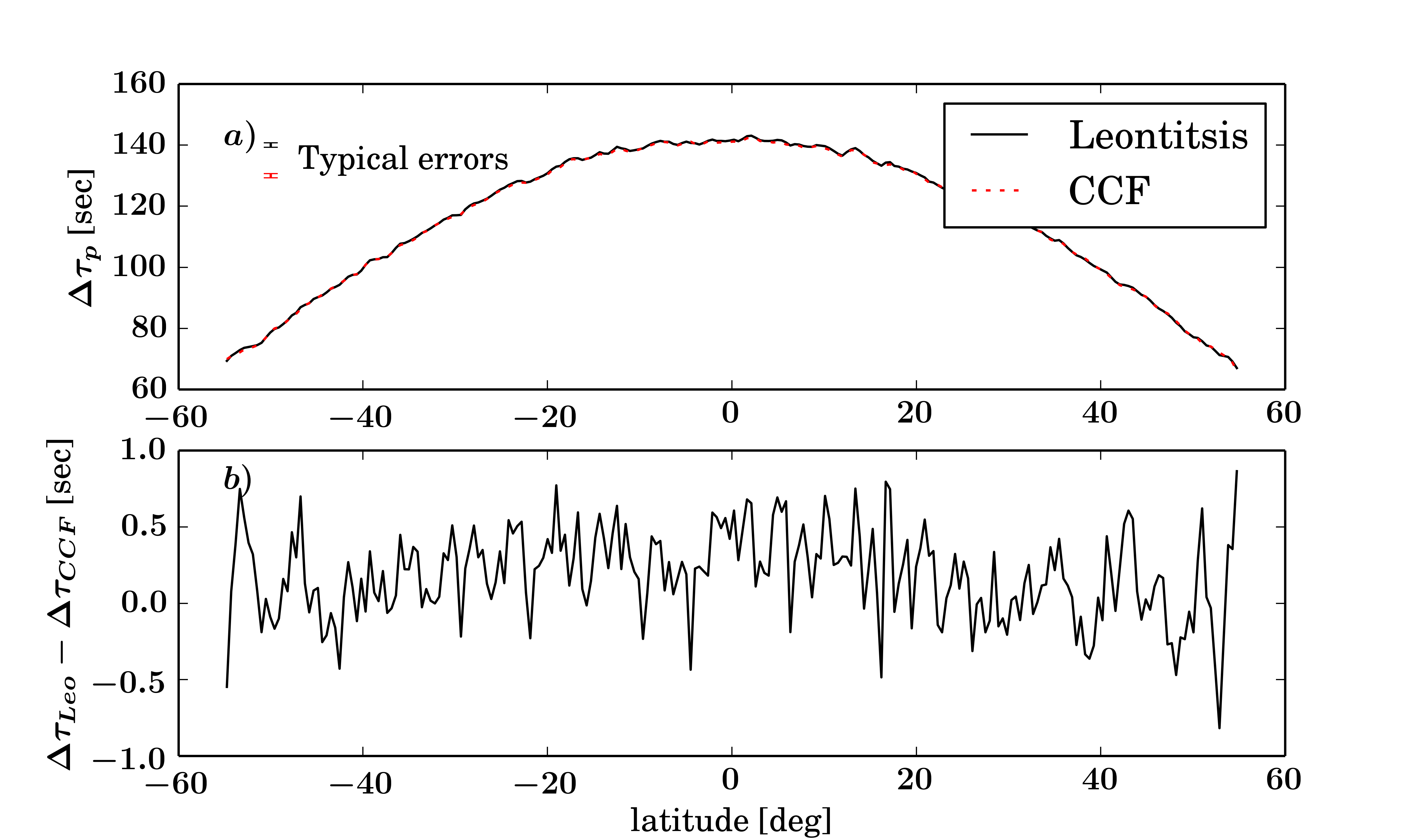}
\label{diffrot}
\caption{Latitudinal profile of solar rotation derived via MI and CCF
methods. (a) The deviation in phase travel time [$\Delta \tau_p$]
between east-going and west-going waves over a range of latitudes,
using the mutual average information algorithm of Leontitsis and using
the traditional CCF approach. (b) Difference between the phase travel-time
deviations as calculated using the approach of Leontitsis, $\Delta
\tau_{Leo}$, and using the CCF method, $\Delta \tau_{CCF}$.}
\end{figure}

Using the histogram method of \inlinecite{leontitsis2001}, we were able to quickly
construct a differential rotation profile.  In this case, rather than
use an annulus, we used the section of the annulus which was
longitudinally displaced in the direction of solar rotation. As the
value given in this algorithm depends on the binning procedure, we did not
make use of the amplitude but merely fit a cosine--squared wavelet and looked for
deviations in the phase travel-time difference between east-going and
west-going waves. \edit{Each step in the analysis is as follows
\begin{enumerate}
\item For each point, get a time series $X_t$ from the data cube.
\item Calculate the average time series for an arc displaced in the direction of
solar rotation, $Y_t$.
\item For each $\tau$ calculate the probability densities using histograms for 
$X_t$, $Y_{t+\tau}$ and the joint signals $(X_t,Y_{t+\tau})$ and $(X_{t+\tau},Y_t)$
for east-traveling and west-traveling waves respectively.
\item Using these probability densities calculate $\hat{I}(\tau)$ from Equation
    (\ref{midef}) and calculate $C(\tau)$ from Equation (\ref{ccfdef}).
\item Average $\hat{I}(\tau)$ and $C(\tau)$ over longitude to get a series for 
each latitude.
\item Extract the phase times, $\tau_p$ from the fit of Equation (\ref{cos2}). And
subtract the east-traveling phase time and west-traveling phase time to see the
effect of solar rotation.
\end{enumerate}
}
In Figure 1, we show the difference in the phase
travel-time for a range of latitudes. The data sets are nearly on top
of each other so we do not show the error bars, but give a typical error
bar in the upper-left corner, which represent errors of about 0.75
seconds for MI and 0.68 seconds for the CCF. There is a slight,
systematic difference between the two sets which tends to make
Leontitsis's method give a very slightly larger value than the CCF, a
difference which grows to about $0.25$ seconds at low latitudes and decreases
for larger latitudes.

\subsection{Sunspot Travel-Time Deviations}
\begin{figure}[t]
\centering
\includegraphics[width=\textwidth]{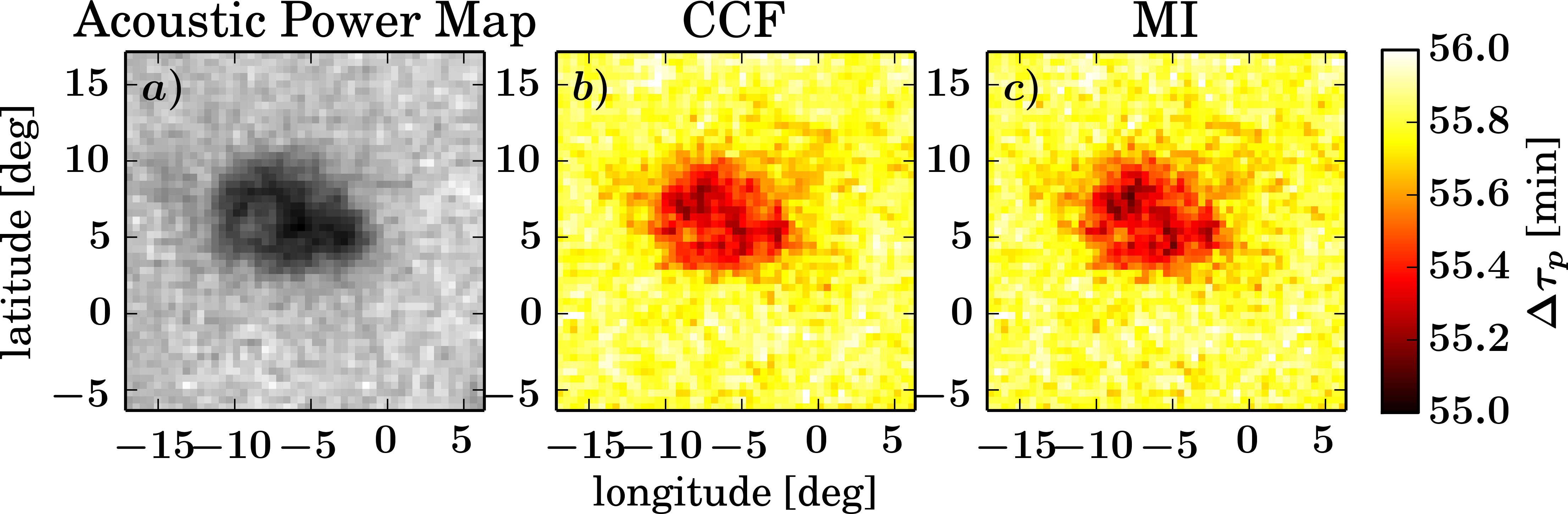}
\caption{(a) A map \edit{of acoustic power} showing the structure 
of AR\,10484 as a comparison to the travel-time maps.  (b) Travel-time 
deviation map calculated using the CCF method. (c) Travel-time deviation map
calculated using the $k$-nearest neighbor MI method.}
\label{spotfig}
\end{figure}
\begin{figure}[h]
\centering
\includegraphics[width=\textwidth]{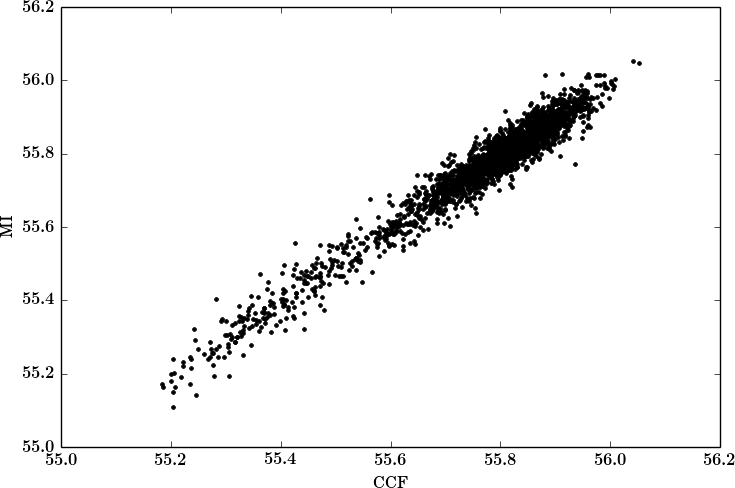}
\caption{\edit{Scatter plot of travel-time deviations (in units of minutes) due to AR\,10484 
for MI and CCF.}}
\label{scatter}
\end{figure}

Using the $k$-nearest neighbor approach we generated a phase travel-time
deviation map of an active region.  Figure \ref{spotfig} shows a phase travel-time
map for AR\,10484, on 23 October 2003, as
well as an acoustic power map for the day as a
comparison. The active region is seen as a decrease in phase travel-time
for outward--traveling waves, by about 20\,--\,40 seconds -- a result
consistent with previous findings \citep{kholikov2004}.  The annuli were
chosen with radii of $7.54^\circ$, $8.47^\circ$, and $9.41^\circ$, so
that there was minimal overlap between the active region and the
annuli. \edit{ The analysis went as follows
\begin{enumerate}
\item Generate data cube for tracked sunspot.
\item For each point, get time series $X_t$ and averaged time series $Y_t$
for the annulus at the three distances.
\item Calculate $C(\tau)$ and fit Gabor wavelet and extract phase time.
\item (MI) Generate cartesian product for a given $\tau$, $Z = (X_t,Y_{t+\tau})$.
\item (MI) Sort data by first dimension. For each point [$(x_i,y_i)$], check neighbors 
and keep list of $k$ neighbors sorted under the maximum norm. When the distance 
from the point to its next neighbor is bigger than the $k$th neighbor under the 
maximum norm, the $k$th nearest neighbor is found. Get distance $d_k(x_i,y_i)$.
\item (MI) For each point, count how many points fall in the 1D neighborhoods of size 
$d_k(x_i,y_i)$, call it $n_x(x_i,y_i)$ for the first dimension and $n_y(x_i,y_i)$
for the second dimension.
\item (MI) Calculate $\psi(n_x(x_i,y_i)+1)$ and $\psi(n_y(x_i,y_i)+1)$ for each 
point and average values over the entire time series.
\item (MI) Plug into Equation (\ref{micalc}) to get $\hat{I}(\tau)$. Repeat for each $\tau$.
\item (MI) Fit Equation (\ref{cos2}) to $\hat{I}(\tau)$ series and extract phase time
\item Average phase time for three distances.
\end{enumerate}
}
 Other parameters are affected by the presence of the active
region, such as group travel time; however, none of the other
parameters provide the level of definition that the phase travel-time
map offers. The extent of the region is fully covered by the travel-time
map, and to a lesser extent so are some of the interior and edge
structures. \edit{The values given by the CCF and MI methods are in good
agreement as can be seen in Figure \ref{scatter}, where the points are
distributed closely around the diagonal.}

\subsection{Time--Distance Diagram}
\begin{figure}[h!]
\centering
\includegraphics[width=\textwidth]{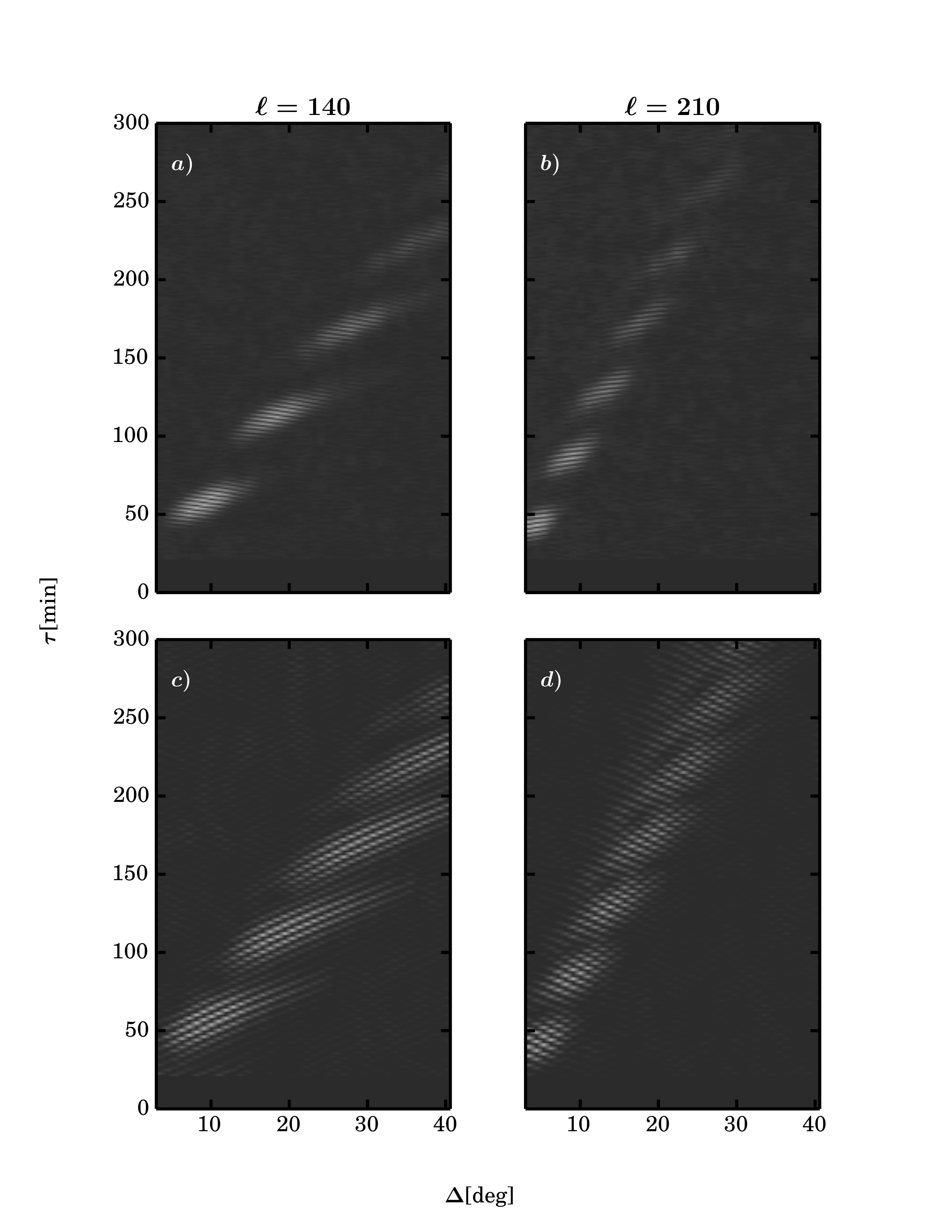}
\label{tdfig}
\caption{\edit{(a) MI time--distance diagram for $\ell = 140$, (b) MI
time--distance diagram for $\ell=210$, (c) CCF time--distance diagram for 
$\ell=140$, (d) CCF time--distance diagram for $\ell=210$.}}
\end{figure}

\begin{figure}[h!]
\centering
\includegraphics[width=\textwidth]{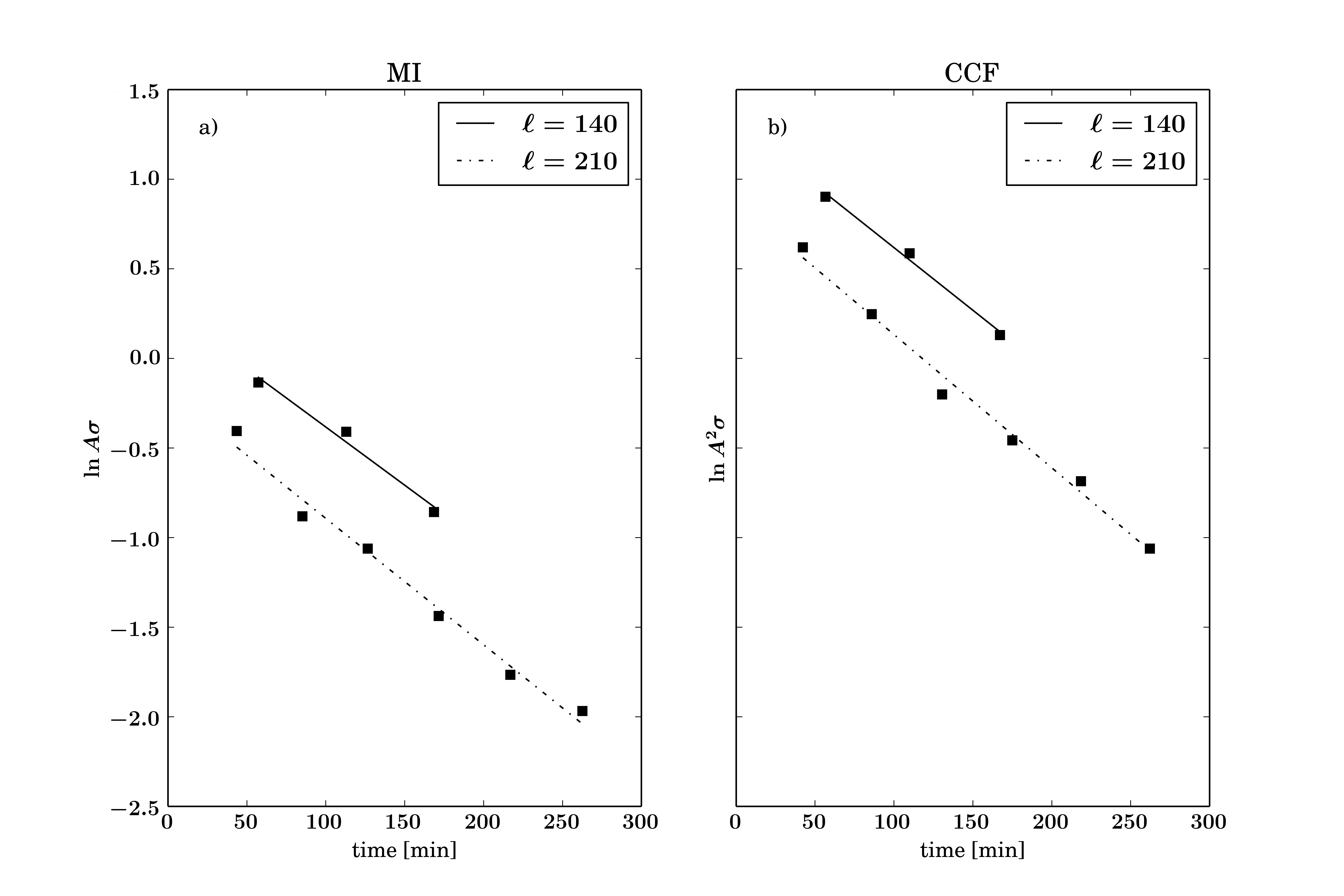}
\label{lifetime}
\caption{(a) Logarithm of amplitude of MI \edit{times the width} \textit{vs.} 
time where the amplitudes \edit{and widths are taken at the maximums of 
each skip}. (b) Logarithm of amplitude squared of CCF times the width
\textit{vs.} time. The slopes are
the negative of the inverse of the lifetime. For MI they are \edit{
$T_{MI,140} = 154.02$ min and $T_{MI,210} = 141.82$ min. The CCF gives
$T_{CCF,140} = 143.03$ min $T_{CCF,210} = 134.48$ min. The errors are on
the order of about $0.6$ sec.}}
\end{figure}

We generated a time--distance diagram, much like the CCF would produce,
using the $k$-nearest neighbor algorithm.  This data was constructed from
three consecutive daily observations [4320 minutes]. The quiet region
near the disk center was used to avoid any travel-time perturbations
due to active regions.  A central point was picked and the MI was
computed with respect to annuli over varying distances.  Then the
calculations were repeated for a different central point, and
again for a $10\times 10$ central square and the results were averaged,
producing the diagrams in Figure 3. \edit{ The analysis went as follows
\begin{enumerate}
\item Pick a central point on the Sun and get a time series $X_t$.
\item Starting at a distance of $3^\circ$, calculate the average time
series, $Y_t$, for an annulus of width $0.94^\circ$.
\item (CCF) Calculate $C(\tau)$ for a range of $\tau$ from 20\,--\,300.
\item (MI) Calculate $\hat{I}(\tau)$ using the same process as the sunspot
for the same range of $\tau$ as the CCF.
\item Repeat for distances increasing by an increment of $0.94^\circ$.
\item Repeat for different central point in the $10\times 10$ central square.
\item Average over central points.
\end{enumerate}}
 The left side shows data which have
been derived from filtered SH time series centered at $\ell=140$, while
the right side is from filtered SH time series centered at $\ell=210$.
\edit{The relation which takes into account dispersion and dissipation for the 
CCF is}

\begin{equation}
\edit{\left(A^2\sigma\right)(t) = A^2_0\sigma_0 \mathrm{e}^{-t/T}},
\end{equation}
\edit{
where $T$ is the lifetime of the wave. We expect then that the MI would
approximately follow this relation with $A^2$ replaced by $A$, a constant change
in $\sigma$ would not affect the lifetime}.
Figure 4 shows the \edit{respective quantities for MI and CCF for} each skip 
on a logarithmic scale against the time of the
maximum amplitude [the group time [$\tau_g$]] where the slope of the
resulting line is $-T^{-1}$. To obtain the amplitudes, we fit Equation
(\ref{cos2}) to the MI data for the time around the maximum of the
skip. This is the same as the method we used to get the CCF data, except
that in that case a Gabor wavelet was used. The error in the amplitude
is of the order of $10^{-6}-10^{-7}$ for MI and $10^{-5}$ for CCF \edit{
and the error in the width is of the order of $10^{-3}-10^{-2}$ for MI and
$10^{-2}-10^{-1}$ for the CCF. Both methods give a similar result for the 
lifetime, with MI giving an only slightly longer lifetime. For MI we have
the lifetimes $T_{MI,140}=154.02$ min and $T_{MI,210}=141.82$ min, and for the CCF
we have $T_{CCF,140}=143.03$ min and $T_{CCF,210}=134.48$ with errors in the range
of $0.6$ sec. Both show an increase in the lifetime of the longer wavelength wave
$\ell = 140$ wave as expected; however, the MI lifetimes are noticeably longer
than the CCF lifetimes.}

\section{Discussion}
\label{sec:discuss}

In Sections 5.1 and 5.2 we showed that MI, and MI-like, measures can be
used to reproduce some of the well-known results of helioseismology
based on the CCF, and in Section 5.3 we showed that MI can produce a
time--distance diagram which can be used to calculate the rate of
information loss and wave lifetimes. There has been work done using
the CCF to calculate wave lifetimes
\citep[\emph{e.g.}][]{chou2001,burtseva2007}, but in order to ensure
consistency we calculated our own lifetimes for the CCF. 
\edit{We found that MI gives a slightly longer lifetime, suggesting that the
lifetime of linear correlations which the CCF quantifies is slightly 
shorter than the lifetime when taking into account nonlinear correlations.
Both MI and CCF values fall} in line with other results, such as that of
\citet{chen1996}, which gives a lifetime of an $\ell = 205$ wave to be
about 1.5\,--\,3 hours, based on the absorption of the waves in sunspots.

\edit{The CCF is easier to work with than MI, but MI brings a more physical
and complete picture.} We see that for phase--time calculations \edit{the two
methods are in excellent agreement. For the phase travel time deviations
due to AR\,10484, the values are in very good agreement} \edit{(see, Figure \ref{scatter})}. 
\edit{ In the sunspot,} the phase time deviation value is about 
$-$40 seconds on the inside and $-$20 seconds on the edge of the spot. \edit{These} phase 
time deviation values may be interpreted as a result of outward \edit{ (Evershed)} flows, from the 
point inside the sunspot to an annulus outside. \edit{But when the amplitudes are
used, the two methods give slightly different answers, which is to be expected
as the quantities are capturing different qualities of the wave. The fact
that the MI lifetime is longer means that nonlinear correlations persist 
longer than linear correlations.}.

There has been some work in tracking information flow and its
relationship to thermodynamic properties for very simple systems
\citep[\emph{e.g.}][]{sagawa2012,barato2013:1,barato2013:2,horowitz2014}.
These authors attempt to derive the relationship between
information theoretic entropy flow and thermodynamic information flow.
In equilibrium situations the two quantities are one and the same
\citep{jaynes1957}, but for nonequilibrium situations their
relationship will depend on the system under consideration. As this is
a young field, there has not been much work on systems where the states 
of the system are drawn from a continuum, as are velocity measurements, 
and not a discrete space; however, one can already see the beginnings of
a framework in which one can talk about the energetics of information flow.

\edit{In this work, we generate a phase--time differential-rotation profile, calculate the 
phase travel time deviation due to a sunspot, and calculate the 
rate of information loss as a wavepacket travels through the chaotic solar atmosphere,
representing the lifetime of the wavepacket. For the differential-rotation 
profile we tested the algorithm of \citet{leontitsis2001}.}
The $k$-nearest neighbor algorithm was tested on the sunspot data, and then 
again to generate a time--distance diagram like the CCF method would produce,
from which we derive the wavepacket lifetimes.
It must be kept in mind that the MI values given here are the observed
MI values, not the physical MI values for the Sun. The data that we used have
a cadence of one minute and a resolution of 0.47 degrees per pixel, so
that the velocity field is coarse-grained even before we introduce
annuli.  This is the same as applying a discretization and averaging filter to
the physical data. To get more realistic values of MI would require a cadence and
resolution smaller than the dynamic ranges of the Sun. This needs to be 
explored using data from instruments such as the 
\textit{Solar Dynamic Observatory/Helioseismic and Magnetic Imager} (SDO/HMI) and the upcoming
\textit{Daniel K. Inouye Solar Telescope} (DKIST, formerly the \textit{Advanced
Technology Solar Telescope}, ATST).

This work represents a proof-of-concept approach to MI. \edit{ The aim was to demostrate
that the new method (MI) can successfully reproduce the known results} \edit{of helioseismology
found with the CCF, as well as provide a different physical perspective.}
There are many avenues for future research. The first, most obvious, project 
would be a systematic study of information loss for a wide range of wavelengths
and distance scales, paying attention to all the parameters of the 
wavepacket, as the amplitude and group time were
being affected by the sunspot. In stochastic modeling, two-point correlation
functions are used as a measure of the correlation of random variables. 
This provides insight only to the level of \textit{linear} correlation
in the random variables.  MI is a better measure because a value of zero is 
a definite sign of independence, whereas one would have to look at all of the
higher moments to arrive at the same conclusion using correlation functions.
MI is derived from a more general view of correlations which allows it to account
for nonlinear relationships.

\edit{Though MI is relatively simple to use computationally, 
it is more difficult to use analytically as the joint
probability density function must be found, using either a stochastic
model or from basic assumptions about the system. This kind of
analysis would be a good place to start future research.}
These are just a few of the ways information theory can be applied in solar
physics.

\begin{acks}
NSO is operated by the Association of Universities for Research in
Astronomy (AURA, Inc.), under a cooperative agreement with the National
Science Foundation (NSF). The SOHO/MDI data used here are 
provided by the SOHO/MDI consortium.
SOHO is a project of international cooperation between ESA and NASA.
\edit{ The authors thank anonymous referee for his/her suggestions that 
allowed the authors to improve the article.}

\end{acks}
\bibliographystyle{spr-mp-sola}
\bibliography{MI_Helioseismology110414.bib}

\begin{thebibliography}{19}
% BibTex style file: spr-mp-sola.bst (nameyear), 2014-02-13
\ifx\bisbn     \undefined \def\bisbn  #1{ISBN #1}\fi
\ifx\binits    \undefined \def\binits#1{#1}\fi
\ifx\bauthor   \undefined \def\bauthor#1{#1}\fi
\ifx\batitle   \undefined \def\batitle#1{#1}\fi
\ifx\bjtitle   \undefined \def\bjtitle#1{\textit{#1}}\fi
\ifx\bvolume   \undefined \def\bvolume#1{\textbf{#1}}\fi
\ifx\byear     \undefined \def\byear#1{#1}\fi
\ifx\bissue    \undefined \def\bissue#1{#1}\fi
\ifx\bfpage    \undefined \def\bfpage#1{#1}\fi
\ifx\blpage    \undefined \def\blpage #1{#1}\fi
\ifx\burl      \undefined \def\burl#1{\textsf{#1}}\fi
\ifx\href      \undefined \def\href#1#2{\textsf{#2}}\fi
\ifx\betal     \undefined \def\betal{\textit{et al.}}\fi
\ifx\bctitle   \undefined \def\bctitle#1{#1}\fi
\ifx\beditor   \undefined \def\beditor#1{#1}\fi
\ifx\bbtitle   \undefined \def\bbtitle#1{\textit{#1}}\fi
\ifx\bedition  \undefined \def\bedition#1{#1}\fi
\ifx\bseriesno \undefined \def\bseriesno#1{\textbf{#1}}\fi
\ifx\blocation \undefined \def\blocation#1{#1}\fi
\ifx\bsertitle \undefined \def\bsertitle#1{\textit{#1}}\fi
\ifx\bsnm      \undefined \def\bsnm#1{#1}\fi
\ifx\bsuffix   \undefined \def\bsuffix#1{#1}\fi
\ifx\bparticle \undefined \def\bparticle#1{#1}\fi
\ifx\barticle  \undefined \def\barticle#1{}\fi
\ifx\binstitute  \undefined \def\binstitute#1{#1}\fi
\ifx\bpublisher  \undefined \def\bpublisher#1{#1}\fi
\ifx\doiurl    \undefined
  \def\doiurl#1{\href{http://dx.doi.org/#1}{\textsf{DOI}}}\fi
\ifx\arxivurl  \undefined
  \def\arxivurl#1{\href{http://arxiv.org/abs/#1}{\textsf{arXiv}}}\fi
\ifx\adsurl    \undefined
  \def\adsurl#1{\href{http://adsabs.harvard.edu/abs/#1}{\textsf{ADS}}}\fi
\ifx\botherref \undefined \def\botherref#1{}\fi
\ifx\url       \undefined \def\url#1{\textsf{#1}}\fi
\ifx\bchapter  \undefined \def\bchapter#1{}\fi
\ifx\bbook     \undefined \def\bbook#1{}\fi
\ifx\bcomment  \undefined \def\bcomment#1{#1}\fi
\ifx\oauthor   \undefined \def\oauthor#1{#1}\fi
\ifx\citeauthoryear \undefined\def \citeauthoryear#1{#1}\fi
\def\endbibitem {}
\ifx\bconflocation  \undefined \def\bconflocation#1{#1} \fi

\bibitem[\protect\citeauthoryear{{Barato}, {Hartich}, and
  {Seifert}}{2013a}]{barato2013:1}
\begin{barticle}
\bauthor{\bsnm{{Barato}}, \binits{A.C.}},
\bauthor{\bsnm{{Hartich}}, \binits{D.}},
\bauthor{\bsnm{{Seifert}}, \binits{U.}}:
\byear{2013}a,
\batitle{{Information-theoretic versus thermodynamic entropy production in
  autonomous sensory networks}}.
\bjtitle{\pre}
\bvolume{87}(\bissue{4}),
\bfpage{042104}.
\doiurl{10.1103/PhysRevE.87.042104}.
\end{barticle}
\endbibitem

\bibitem[\protect\citeauthoryear{{Barato}, {Hartich}, and
  {Seifert}}{2013b}]{barato2013:2}
\begin{barticle}
\bauthor{\bsnm{{Barato}}, \binits{A.C.}},
\bauthor{\bsnm{{Hartich}}, \binits{D.}},
\bauthor{\bsnm{{Seifert}}, \binits{U.}}:
\byear{2013}b,
\batitle{{Rate of Mutual Information Between Coarse-Grained Non-Markovian
  Variables}}.
\bjtitle{Journal of Statistical Physics}
\bvolume{153},
\bfpage{460}.
\doiurl{10.1007/s10955-013-0834-5}.
\end{barticle}
\endbibitem

\bibitem[\protect\citeauthoryear{{Burtseva}
  \textit{et~al.}}{2007}]{burtseva2007}
\begin{barticle}
\bauthor{\bsnm{{Burtseva}}, \binits{O.}},
\bauthor{\bsnm{{Kholikov}}, \binits{S.}},
\bauthor{\bsnm{{Serebryanskiy}}, \binits{A.}},
\bauthor{\bsnm{{Chou}}, \binits{D.-Y.}}:
\byear{2007},
\batitle{{Effects of Dispersion of Wave Packets in the Determination of
  Lifetimes of High-Degree Solar p Modes from Time Distance Analysis: TON
  Data}}.
\bjtitle{\solphys}
\bvolume{241},
\bfpage{17}.
\doiurl{10.1007/s11207-007-0325-4}.
\end{barticle}
\endbibitem

\bibitem[\protect\citeauthoryear{{Chen}, {Chou}, and {TON
  Team}}{1996}]{chen1996}
\begin{barticle}
\bauthor{\bsnm{{Chen}}, \binits{K.-R.}},
\bauthor{\bsnm{{Chou}}, \binits{D.-Y.}},
\bauthor{\bsnm{{TON Team}}}:
\byear{1996},
\batitle{{Determination of the Lifetime of High-l Solar p-Modes from the
  Interaction of p-Mode Waves with Sunspots}}.
\bjtitle{\apj}
\bvolume{465},
\bfpage{985}.
\doiurl{10.1086/177484}.
\end{barticle}
\endbibitem

\bibitem[\protect\citeauthoryear{{Chou} \textit{et~al.}}{2001}]{chou2001}
\begin{barticle}
\bauthor{\bsnm{{Chou}}, \binits{D.-Y.}},
\bauthor{\bsnm{{Serebryanskiy}}, \binits{A.}},
\bauthor{\bsnm{{Ye}}, \binits{Y.-J.}},
\bauthor{\bsnm{{Dai}}, \binits{D.-C.}},
\bauthor{\bsnm{{Khalikov}}, \binits{S.}}:
\byear{2001},
\batitle{{Lifetimes of High-l Solar p-Modes from Time-Distance Analysis}}.
\bjtitle{\apjl}
\bvolume{554},
\bfpage{L229}.
\doiurl{10.1086/321713}.
\end{barticle}
\endbibitem

\bibitem[\protect\citeauthoryear{{Cover} and {Thomas}}{2006}]{coverthomas}
\begin{bbook}
\bauthor{\bsnm{{Cover}}, \binits{T.M.}},
\bauthor{\bsnm{{Thomas}}, \binits{J.A.}}:
\byear{2006},
\bbtitle{{Elements of information theory}},
\bedition{2}nd edn.
\bpublisher{Wiley-Interscience},
\blocation{Hoboken, NJ},
\bfpage{20}.
\end{bbook}
\endbibitem

\bibitem[\protect\citeauthoryear{{Duvall} \textit{et~al.}}{1993}]{duvall1993}
\begin{barticle}
\bauthor{\bsnm{{Duvall}}, \binits{T.L.} \bsuffix{Jr.}},
\bauthor{\bsnm{{Jefferies}}, \binits{S.M.}},
\bauthor{\bsnm{{Harvey}}, \binits{J.W.}},
\bauthor{\bsnm{{Pomerantz}}, \binits{M.A.}}:
\byear{1993},
\batitle{{Time-distance helioseismology}}.
\bjtitle{\nat}
\bvolume{362},
\bfpage{430}.
\doiurl{10.1038/362430a0}.
\end{barticle}
\endbibitem

\bibitem[\protect\citeauthoryear{{Horowitz} and
  {Esposito}}{2014}]{horowitz2014}
\begin{barticle}
\bauthor{\bsnm{{Horowitz}}, \binits{J.M.}},
\bauthor{\bsnm{{Esposito}}, \binits{M.}}:
\byear{2014},
\batitle{{Thermodynamics with Continuous Information Flow}}.
\bjtitle{Physical Review X}
\bvolume{4}(\bissue{3}),
\bfpage{031015}.
\doiurl{10.1103/PhysRevX.4.031015}.
\end{barticle}
\endbibitem

\bibitem[\protect\citeauthoryear{Jaynes}{1957}]{jaynes1957}
\begin{barticle}
\bauthor{\bsnm{Jaynes}, \binits{E.T.}}:
\byear{1957},
\batitle{Information theory and statistical mechanics}.
\bjtitle{Physical Review}
\bvolume{106(4)},
\bfpage{620}.
\doiurl{10.1103/PhysRev.106.620}.
\end{barticle}
\endbibitem

\bibitem[\protect\citeauthoryear{{Kholikov}}{2004}]{kholikov2004}
\begin{bchapter}
\bauthor{\bsnm{{Kholikov}}, \binits{S.}}:
\byear{2004},
\bctitle{{Travel Time Measurements in Sunspots}}.
In: \beditor{\bsnm{{Danesy}}, \binits{D.}} (ed.)
\bbtitle{SOHO 14 Helio- and Asteroseismology: Towards a Golden Future},
\bsertitle{ESA Special Publication}
\bseriesno{559},
\bfpage{513}.
\end{bchapter}
\endbibitem

\bibitem[\protect\citeauthoryear{{Kholikov} and
  {Hill}}{2014}]{kholikovhill2014}
\begin{barticle}
\bauthor{\bsnm{{Kholikov}}, \binits{S.}},
\bauthor{\bsnm{{Hill}}, \binits{F.}}:
\byear{2014},
\batitle{{Meridional-Flow Measurements from Global Oscillation Network Group
  Data}}.
\bjtitle{\solphys}
\bvolume{289},
\bfpage{1077}.
\doiurl{10.1007/s11207-013-0394-5}.
\end{barticle}
\endbibitem

\bibitem[\protect\citeauthoryear{{Kholikov}, {Serebryanskiy}, and
  {Jackiewicz}}{2014}]{kholikov2014}
\begin{barticle}
\bauthor{\bsnm{{Kholikov}}, \binits{S.}},
\bauthor{\bsnm{{Serebryanskiy}}, \binits{A.}},
\bauthor{\bsnm{{Jackiewicz}}, \binits{J.}}:
\byear{2014},
\batitle{{Meridional Flow in the Solar Convection Zone. I. Measurements from
  GONG Data}}.
\bjtitle{\apj}
\bvolume{784},
\bfpage{145}.
\doiurl{10.1088/0004-637X/784/2/145}.
\end{barticle}
\endbibitem

\bibitem[\protect\citeauthoryear{{Kozachenko} and {{Leonenko}, N.
  N.}}{1987}]{kozachenko1987}
\begin{barticle}
\bauthor{\bsnm{{Kozachenko}}, \binits{L.F.}},
\bauthor{\bsnm{{{Leonenko}, N. N.}}}:
\byear{1987},
\batitle{Sample estimate of the entropy of a random vector}.
\bjtitle{Probl. Peredachi Inf.}
\bvolume{23},
\bfpage{9}.
\burl{http://mi.mathnet.ru/eng/ppi/v23/i2/p9}.
\end{barticle}
\endbibitem

\bibitem[\protect\citeauthoryear{{Kraskov}, {St{\"o}gbauer}, and
  {Grassberger}}{2004}]{kraskov2004}
\begin{barticle}
\bauthor{\bsnm{{Kraskov}}, \binits{A.}},
\bauthor{\bsnm{{St{\"o}gbauer}}, \binits{H.}},
\bauthor{\bsnm{{Grassberger}}, \binits{P.}}:
\byear{2004},
\batitle{{Estimating mutual information}}.
\bjtitle{\pre}
\bvolume{69}(\bissue{6}),
\bfpage{066138}.
\doiurl{10.1103/PhysRevE.69.066138}.
\end{barticle}
\endbibitem

\bibitem[\protect\citeauthoryear{Leontitsis}{2001}]{leontitsis2001}
\begin{botherref}
\oauthor{\bsnm{Leontitsis}, \binits{A.}}:
2001,
\textit{Mutual average information},
\url{http://www.mathworks.com/matlabcentral/fileexchange/880-mutual-average-information}.
\end{botherref}
\endbibitem

\bibitem[\protect\citeauthoryear{{Sagawa} and {Ueda}}{2012}]{sagawa2012}
\begin{barticle}
\bauthor{\bsnm{{Sagawa}}, \binits{T.}},
\bauthor{\bsnm{{Ueda}}, \binits{M.}}:
\byear{2012},
\batitle{{Fluctuation Theorem with Information Exchange: Role of Correlations
  in Stochastic Thermodynamics}}.
\bjtitle{Physical Review Letters}
\bvolume{109}(\bissue{18}),
\bfpage{180602}.
\doiurl{10.1103/PhysRevLett.109.180602}.
\end{barticle}
\endbibitem

\bibitem[\protect\citeauthoryear{{Scherrer}
  \textit{et~al.}}{1995}]{sherrer1995}
\begin{barticle}
\bauthor{\bsnm{{Scherrer}}, \binits{P.H.}},
\bauthor{\bsnm{{Bogart}}, \binits{R.S.}},
\bauthor{\bsnm{{Bush}}, \binits{R.I.}},
\bauthor{\bsnm{{Hoeksema}}, \binits{J.T.}},
\bauthor{\bsnm{{Kosovichev}}, \binits{A.G.}},
\bauthor{\bsnm{{Schou}}, \binits{J.}},
\bauthor{\bsnm{{Rosenberg}}, \binits{W.}},
\bauthor{\bsnm{{Springer}}, \binits{L.}},
\bauthor{\bsnm{{Tarbell}}, \binits{T.D.}},
\bauthor{\bsnm{{Title}}, \binits{A.}},
\bauthor{\bsnm{{Wolfson}}, \binits{C.J.}},
\bauthor{\bsnm{{Zayer}}, \binits{I.}},
\bauthor{\bsnm{{MDI Engineering Team}}}:
\byear{1995},
\batitle{{The Solar Oscillations Investigation - Michelson Doppler Imager}}.
\bjtitle{\solphys}
\bvolume{162},
\bfpage{129}.
\doiurl{10.1007/BF00733429}.
\end{barticle}
\endbibitem

\bibitem[\protect\citeauthoryear{Shannon}{1948}]{shannon1948}
\begin{barticle}
\bauthor{\bsnm{Shannon}, \binits{C.}}:
\byear{1948},
\batitle{A mathematical theory of communication}.
\bjtitle{Bell System Technical Journal}
\bvolume{27},
\bfpage{379}.
\end{barticle}
\endbibitem

\bibitem[\protect\citeauthoryear{{Victor}}{2002}]{victor2002}
\begin{barticle}
\bauthor{\bsnm{{Victor}}, \binits{J.D.}}:
\byear{2002},
\batitle{{Binless strategies for estimation of information from neural data}}.
\bjtitle{\pre}
\bvolume{66}(\bissue{5}),
\bfpage{051903}.
\doiurl{10.1103/PhysRevE.66.051903}.
\end{barticle}
\endbibitem

\end{thebibliography}

\end{article} 
\end{document}